\documentclass[usenatbib]{mn2e}

\newcommand{\rxj}{HM~Cnc}
\newcommand{\msun}{\mathrm{M}_\odot}
\newcommand{\rsun}{\mathrm{R}_\odot}
\newcommand{\lsun}{\mathrm{L}_\odot}
\newcommand{\K}{\mathrm{K}}
\newcommand{\kms}{\mathrm{km}\,\mathrm{s}^{-1}}

\usepackage{graphicx}
\title[ULTRACAM Photometry of the ultracompact binaries V407~Vul  and
\rxj]{ULTRACAM Photometry of the ultracompact binaries V407~Vul 
and \rxj}
\author[S.C.C. Barros et al. ]{
  S.C.C. Barros$^{1}$\thanks{E-mail:s.c.barros@warwick.ac.uk}
  T.R. Marsh$^{1}$, V. S. Dhillon$^2$  , P. J. Groot$^{3}$,
  S. Littlefair$^2$,\and G. Nelemans$^3$, G. Roelofs$^3$, D. Steeghs$^{4}$ and  P. J. Wheatley$^{1}$ \\
$^{1}$Department of Physics, University of Warwick, Coventry, CV4
7AL,UK\\
$^{2}$Department of Physics and Astronomy, University of
Sheffield,Sheffield , S3 7RH, UK\\
$^{3}$Department of Astrophysics/IMAPP, Radboud University Nijmegen, P.O. Box 9010, 
NL - 6500 GL Nijmegen, The Netherlands\\
$^{4}$Harvard-Smithsonian Center for Astrophysics, Cambridge, USA }

\begin{document}
\date{Accepted ; in original form 2005 August 10}

%\baselinestretch{2}

\pagerange{\pageref{firstpage}--\pageref{lastpage}} \pubyear{2002}

\maketitle

\label{firstpage}

\begin{abstract}
  V407~Vul (RXJ1914.4+2456) and \rxj\ (RXJ0806.3+1527) are X-ray
  emitting stars with X-ray light curves that are 100\% modulated on
  periods of 569 and 321 seconds respectively. These periods are
  thought possibly to represent the orbital periods of close pairs of
  white dwarfs. In this paper we present optical light curves taken
  with the high-speed CCD camera ULTRACAM on the 4.2m WHT in May 2003
  and August 2005 and with the VLT in November 2005.  The optical and
  X-ray light curves of \rxj\ have been reported as being in
  anti-phase, but we find that in fact the X-rays peak around 0.2
  cycles after the maximum of the optical light, as seen also in
  V407~Vul. The X-ray/optical phase shifts are well explained under
  the accreting models of the systems if most of the optical
  modulation comes from the heated faces of the mass donors and if the
  X-ray emitting spots are positioned in advance of the mass donors,
  as is expected given the angular momentum of the accreting material.
  Some optical emission may also come from the vicinity of the X-ray
  spot, and we further show that this can explain the non-sinusoidal
  lightcurves of \rxj. On the basis of this model we constrain the
  temperature of the heated face of the companion star finding a
  bolometric luminosity $> 10^{33} \, \mathrm{ergs}\,\mathrm{s}^{-1} $
  and a distance, $d > 1.1\,$kpc. We can identify no explanation for
  the X-ray/optical phase-shifts under the intermediate polar and
  unipolar inductor models of the systems. The only significant
  difference between the two stars is that V407~Vul is observed to
  have the spectrum of a G star. The variation in position on the sky
  of a blend of a variable and a constant star can be used as a
  measure of their separation, and is sensitive to values well below
  the limit set by seeing. We apply this "pulsation astrometry" to
  deduce that the G star is separated from the variable by about
  $0.027''$ and hence plays no role in the variability of V407~Vul. We
  show that light travel time variations could influence the period
  change in V407~Vul if it forms a triple system with the G star.
\end{abstract}

\begin{keywords}
binaries: close-- stars: individual: V407~Vul, \rxj\ -- white dwarfs --
stars: magnetic fields -- X-rays: stars -- astrometry
\end{keywords}

\section{Introduction}

V407~Vul \citep[RXJ1914.4+2456][]{motch1996a} and \rxj\
\citep[RXJ0806.3+1527,][]{israel1999a,burwitz2001a} were both
discovered in the ROSAT all sky survey and have very similar X-ray
properties. They have periods of $P = 569$ sec
\citep{cropper1998a,motch1996a} and $P =321$ sec \citep{israel1999a}
respectively.  In each star, only one period (and its harmonics) has
been observed \citep{ramsay2000a,ramsay2002b,israel02a} at all
wavelengths.  Taken together, the observations have lead to a belief
that the periods may be orbital, making these the shortest period
binary stars known, and probably composed of pairs of white dwarfs.
This would make these systems strong emitters of gravitational waves
and possible progenitors or representatives of semi-detached AM~CVn
stars.

There are several rival models for these stars, all of them based upon binary
systems.  The intermediate polar (IP) model \citep{motch1996a,israel1999a, norton04a}
is the only one in which the pulsation periods are not assumed to be orbital.
In this model, the pulsations are ascribed to the spin of white dwarfs accreting
from non-degenerate secondary stars; the orbital periods are presumed
undetectable. The other three models all invoke double white dwarf binaries in
which the pulsation periods are the orbital periods. There is one detached model
(i.e non-accreting), the unipolar inductor model \citep{wu02a}, also called the
electric star model because it is powered by the dissipation of electric
currents induced by an asynchronism between the spin period of a magnetic white
dwarf and the orbital period within a detached double white dwarf binary.  The
other two models each employ semi-detached, accreting double white dwarfs: one
is magnetic, the double degenerate polar model 
\citep{cropper1998a,ramsay2002b,israel02a}, while the other is non-magnetic, the
direct impact model \citep{nelemans2001b,marsh2002a,ramsay2002a}, in which, due
to the compact dimensions of these systems, the mass transfer streams crash
directly onto the accreting white dwarfs.

It has proved hard to decide which, if any, of the models is correct.
Compared to typical accreting systems, \rxj\ has a weak optical line emission, while V407 Vul has none at all. This favours the unipolar inductor
model which is the only one without accretion. The unipolar inductor model, along
with the IP model, is also favoured by the observed decrease in pulsation
periods \citep{strohmayer2002a, strohmayer04a, hakala2003a, strohmayer03a,
  hakala2004a} although recently accreting models with long-lasting spin-up
phases have been developed \citep{DAntona2006,deloye2006}. The shapes and phases of
the X-ray light curves on the other hand count against the unipolar inductor model
\citep{susana2005} which can only accommodate the high X-ray luminosity
of V407~Vul with a white dwarf that spins faster than its orbit
\citep{marsh2005a,dallosso2006a,dallosso2006b}. The accreting double-degenerate models on the
other hand lead to high accretion rates and strong heating of the white dwarf,
particularly in the case of \rxj, which is required to be at a distance of
$4$ to $20\, \mathrm{kpc}$, and well out of the Galactic plane \citep{bildsten2005,DAntona2006}.  At
the moment therefore, there is no clear winner, or even leading contender
amongst the models and better observational constraints are a priority.

Previous studies of the systems have focused mainly upon the properties of the
X-ray light curves with optical data used mainly to track the decreasing periods
with less attention being paid to the shapes of the light curves.  With the work
of \citet{DAntona2006} and \citet{deloye2006} adding uncertainty to the interpretation of the period change
measurements, the light curves themselves take on more significance.  In this
paper we present high-speed photometry of these systems in three simultaneous
bands taken in the hope of using the optical characteristics to learn more
about the systems.  In section~2 we report our observations and data reduction.
In section~3 we present our results. In section~4 we use our results to try to
determine the origin of the optical pulses and explore the consequences for the accretion geometry in these systems.

\section{Observations and reduction}
\label{sec:observa}

We observed with the high-speed CCD camera ULTRACAM \citep{ultracam2001} mounted
on the $4.2$m William Hershel telescope (WHT) in La Palma on May 2003 and August 2005,
and mounted on the UT3 unit (Melipal) of the Very Large Telescope (VLT) in Chile in November 2005. For
V407~Vul we have observations on five consecutive nights from the $21^{st}$ to
$25^{\mathrm{th}}$ of May, 2003 with a total of approximately 3600 frames of 9.7 sec exposure
in the $i'$, $g'$ and $u'$ filters and another $2000$ frames of 15~sec
exposure in five extra nights from the $27^{\mathrm{th}}$ of August to the $1^\mathrm{st}$ of
September 2005 in $r'$, $g'$ and $u'$.  For \rxj\ we have around
2000 frames taken in four nights from the $21^{\mathrm{st}}$ to $25^{\mathrm{th}}$ of May with
10.1~sec exposures in $i'$, $g'$ and $u'$ and another $18$,$000$ frames taken in
November 2005 in $r'$, $g'$ and $u'$ with exposures of 1 to 6~sec.  The
observing conditions are summarised in Table~\ref{tbl:log1}. All the times were
transformed to TDB, and then shifted to time as observed at the solar system
barycentre using the IDL routine \textit{barycen} and recorded as a modified
Julian day MJD(TDB).
\begin{table}
\centering
\begin{tabular}{||l|l|c|c|c||} \hline  \hline
\textbf{Target} & \textbf{Date} & \textbf{UT}  &
\textbf{Seeing, clouds}  \\ \hline
V407~Vul &21 May 2003&  05:33 - 06:25  & 1.0, clear     \\
V407~Vul &22 May 2003&  03:28 - 04:24  & 1.0, clear    \\
V407~Vul &22 May 2003&  04:54 - 06:25  & 1.0, clear     \\
V407~Vul &23 May 2003&  02:25 - 04:24  & 1.0, clear     \\
V407~Vul &24 May 2003&  02:48 - 03:41  & 1.0, some     \\
V407~Vul &24 May 2003&  04:50 - 06:18  & 1.0, clear     \\
V407~Vul &25 May 2003&  01:45 - 02:29  & 1.2, clear     \\

V407~Vul &25 May 2003&  03:19 - 04:41  & 1.2, clear     \\
V407~Vul &27 Aug 2005&  21:10 - 01:02  & 1.1, clear     \\
V407~Vul &28 Aug 2005&  21:05 - 22:38  & 0.9, clear     \\
V407~Vul &30 Aug 2005&  20:50 - 23:55  & 0.8, dusty     \\
V407~Vul &31 Aug 2005&  20:49 - 22:56  & 0.7, dusty     \\
V407~Vul &01 Sep 2005&  20:45 - 22:58  & 0.9, dusty     \\
\rxj\ &21 May 2003&  22:11 - 23:30  & 1.2, clear     \\
\rxj\ &22 May 2003&  21:54 - 22:57  & 1.0, clear     \\
\rxj\ &23 May 2003&  21:57 - 22:54  & 1.0, clear     \\
\rxj\ &25 May 2003&  21:55 - 22:39  & 1.3, clear     \\
\rxj\ &27 Nov 2005&  05:03 - 06:51  & 1.3, clear     \\
\rxj\ &28 Nov 2005&  05:10 - 08:47  & 1.0, clear     \\
\rxj\ &29 Nov 2005&  05:35 - 08:51  & 0.8, clear     \\ \hline\hline
\end{tabular} \\
\caption{Observation log. \label{tbl:log1}}
\end{table}

The data were reduced using the ULTRACAM pipeline. We tried ``optimal''
photometry \citep{naylor1998}, variable aperture photometry and fixed aperture
photometry to extract the light curves. Optimal photometry gave the higher
signal-to-noise with the only exception the $r'$ band in the August 2005
data, for which we used a fixed aperture radius. Optimal photometry requires
the profiles to be identical in shape and can cause difficulties if this is not the
case and we believe that in this one case this outweighed the improvement in
stochastic noise. The subsequent data analysis was carried out with IDL.
V407~Vul is in a crowded field so care was taken to prevent the sky annulus
from being contaminated by other stars. It is trickier to allow for the faint
stars that can contaminate the target aperture in poor seeing.  These are a
particular problem in the $i'$ filter (May 2003 data) where we found the flux
could increase by as much as $5$\% in the poorest seeing.  Although relatively
few of the data were affected by this, we corrected for it by fitting and
removing the trend of flux versus seeing from the $i'$ data.  Finally, the $g'$
data from the second half of the second run of the $22^{\mathrm{nd}}$ of May 2003 and the
second half of the second run of the $24^{\mathrm{th}}$ of May 2003 could not be used
because V407~Vul was unfortunately positioned close to a column of poor charge
transfer on the $g'$ CCD.

In the May 2003 observations of V407~Vul we used two comparison stars, one (c1)
for the $i'$ and $g'$ bands and the other (c2) for the $u'$ images (because c1
was too faint in $u'$).  The position relative to the target and the magnitudes
of these comparison stars and the one used for \rxj\ are given in
Table~\ref{tbl:comp1}. In the August 2005 observations of V407 Vul we only used
comparison star c2 because c1 was saturated in $r'$ due to the longer exposure
time. This run also suffered from Saharan dust that lead to an extra and
variable extinction of $\sim 0.5$ magnitudes at the zenith making it impossible
to derive an absolute calibration to better then 0.2 magnitudes. Therefore we
used the $g'$ and $u'$ magnitudes of c2 calculated in May 2003 to calibrate the
August 2005 data. To obtain the $r'$ magnitude of c2 we applied the same
correction as for $g'$.

The measured mean magnitudes of the systems are given in Table~\ref{tbl:magni}.
As far as possible, the magnitude calibration was carried out by comparing the
target and the comparison at the same airmass as we did not have sufficiently
long runs to estimate accurate extinction coefficients.  The uncertainties of
the comparison star for \rxj\ are dominated by the uncertainties in the
extinction coefficients for the night both in May 2003 ($i'$) and in August 2005
($r'$, $g'$ and $u'$) because in this case we did not observe the target and the
comparison at exactly the same airmass and some correction was needed.

\begin{table*}
\centering
\begin{tabular}{||l|r@{$\pm$}l|r@{$\pm$}l|r@{$\pm$}l|r@{$\pm$}l||} \hline  \hline
\textbf{Target}  &   \multicolumn{2}{c}{\textbf{$i'$}} &    
\multicolumn{2}{c}{\textbf{$r'$}}&   \multicolumn{2}{c}{\textbf{$g'$}} &   
\multicolumn{2}{c}{\textbf{$u'$}}\\ \hline
V407~Vul May 2003  &18.95 &0.05 &  \multicolumn{2}{c}{-}       &20.30  &0.06  &21.56  &0.10  \\
V407~Vul Aug 2005  &\multicolumn{2}{c}{-} &19.3 &0.1            &20.29  &0.06  &21.53  &0.08 \\
\rxj\    May 2003  &21.5 &0.1 &  \multicolumn{2}{c}{-}         &20.9   &0.1   &20.5   &0.1  \\
\rxj\    Nov 2005  &\multicolumn{2}{c}{-} &21.21 &0.10          &20.77  &0.11  &20.51  &0.12    \\ \hline\hline
\end{tabular} 
\caption{Magnitudes measured for the two targets.\label{tbl:magni} }
\end{table*}

\begin{table*}
\centering
\begin{tabular}{||l|c|c|r@{$\pm$}l|r@{$\pm$}l|r@{$\pm$}l|r@{$\pm$}l||} 
\hline  \hline
\textbf{Comparison} & \textbf{$\Delta \alpha$ arcsec}  &
\textbf{$\Delta \delta$ arcsec} &   \multicolumn{2}{c}{\textbf{$i'$}} &    
\multicolumn{2}{c}{\textbf{$r'$}}&   \multicolumn{2}{c}{\textbf{$g'$}} &   
\multicolumn{2}{c}{\textbf{$u'$}}\\ \hline
V407~Vul c1&  +3.1  &  -8.4  &14.21 &0.01 & \multicolumn{2}{c}{saturated}  &16.26   &0.01  & 
19.78 & 0.03 \\
V407~Vul c2&  +39.5 & -37.0  &15.73 &0.01 &16.08 &0.1  &16.96  &0.01  & 
18.84 & 0.03\\
\rxj\      &  -16.9 & -16.4  &15.25 &0.11 & 15.31     & 0.10    &16.00  
&0.11  & 17.73 & 0.12 \\ \hline\hline
\end{tabular} \\
\caption{Positions relative to the target and magnitude of the comparison 
stars used to flux calibrate the
data.\label{tbl:comp1} }
\end{table*}

\section{Results}

\subsection{Ephemerides} 

\label{sec:efi}
To compare our optical data with the published X-ray data we had to
fold our data, on the X-ray ephemeris. Unfortunately none of the
ephemerides published so far
\citep{strohmayer04a,strohmayer05a,israel2003a,israel2004a,ramsay2006}
give the covariance terms of the fitted coefficients which are needed
for a correct evaluation of the uncertainties.  Therefore we had to
digitise and fit the data of \citet{strohmayer04a,strohmayer05a} and
\citet{ramsay2006} so that we could obtain a timing solution whose
uncertainties we could compare with our data. When we did this we
realised that there was an inconsistency between the ephemerides of
V407~Vul published by \citet{strohmayer04a} and \citet{ramsay2006}.
After investigation we concluded that \citeauthor{strohmayer04a}'s
(\citeyear{strohmayer04a}) ephemeris is probably in error because the
ROSAT times were not corrected from UTC to TT. We therefore use our
fitted \citeauthor{ramsay2006}'s (\citeyear{ramsay2006}) ephemeris
(Table~\ref{tbl:efe1}) for V407~Vul which is similar to ephemeris
given in \citet{ramsay2006} but has a slightly different $\dot{\nu}$.
For \rxj\ we used \citeauthor{strohmayer05a}'s
(\citeyear{strohmayer05a}) ephemeris. Both ephemerides and respective
covariance terms are given in Appendix~A where we provide full
details of our investigations.

\subsection{V407 Vul}

\label{sec:v407vul}

\begin{figure*}
\hspace*{\fill}
\includegraphics[width=\columnwidth]{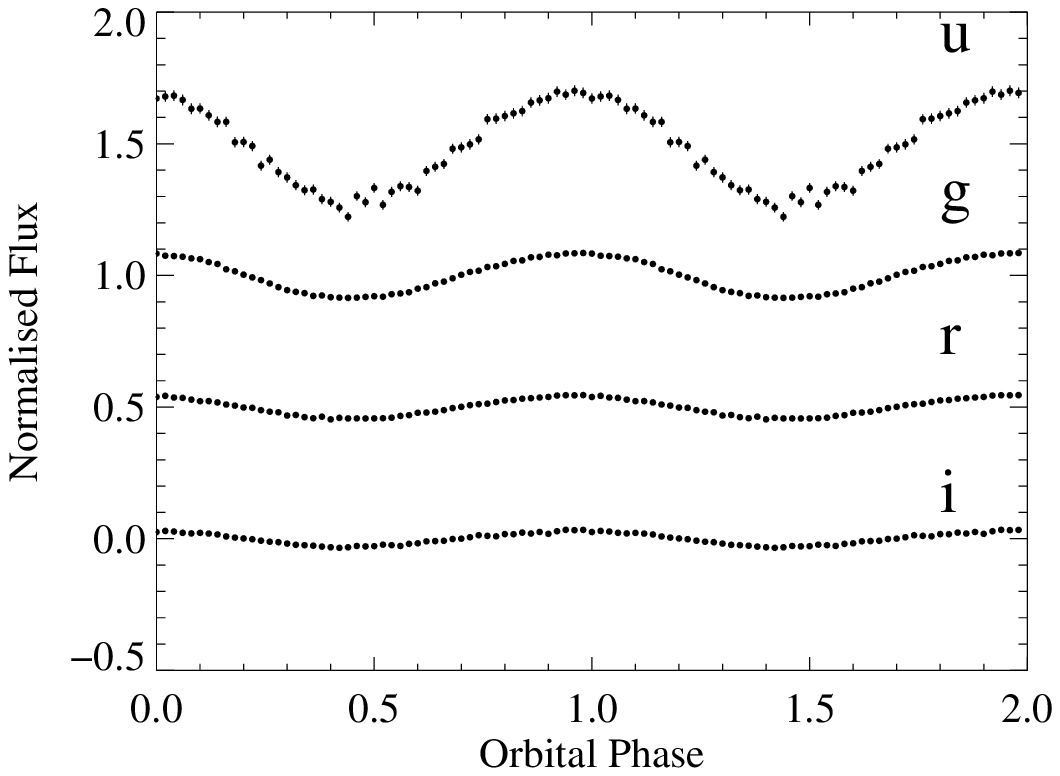}
\hspace*{\fill}
\includegraphics[width=\columnwidth]{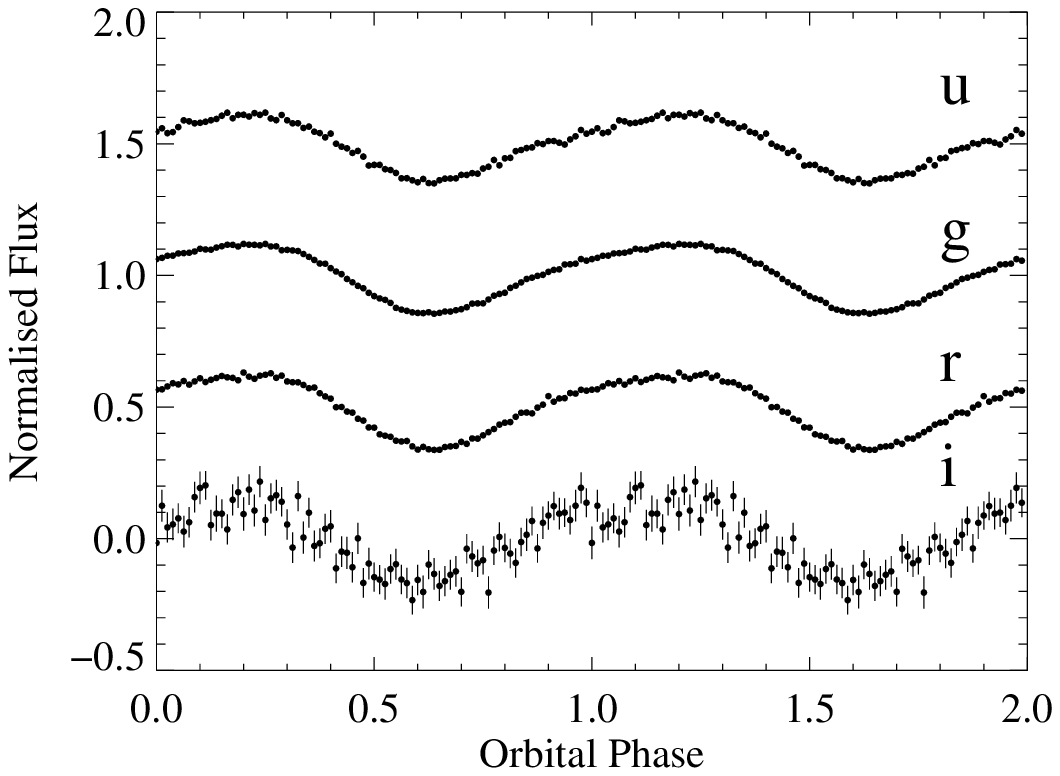}
\hspace*{\fill}
\caption{ Phase-folded light curves of  V407 Vul (left) and \rxj\
(right) using Table~\ref{tbl:efe1} and  \citeauthor{strohmayer05a}'s 
(\citeyear{strohmayer05a}). The flux is normalised 
to unity in each case. The different filters
are displaced vertically for clarity.}
\label{fig:v407lc}
\end{figure*}

We show our phase-folded light curves of V407~Vul folded on the
ephemeris of Table~\ref{tbl:efe1} on the left of
Fig.~\ref{fig:v407lc}. The two datasets (May 2003 and August 2005)
were rebinned to 100 phase bins using inverse-variance weighting to maximise the
signal-to-noise ratio.

We computed the Lomb-Scargle periodogram \citep{lomb1976,scargle1982} and
confirmed the previously measured period of 569 sec. We then tested how close
the signal is to a perfect sinusoid by fitting a sine wave at the fundamental
frequency and at the second and third harmonics. The third harmonic is
consistent with zero. The results for the relative amplitude and the phase at
maximum of the fundamental (i.e. the ``first harmonic'') and the second harmonic
are shown in Table~\ref{tbl:harmo} which also shows the corresponding results
\begin{table*}
\centering
\begin{tabular}{||l|r@{$\pm$}l|r@{$\pm$}l|r@{$\pm$}l|r@{$\pm$}l| 
r@{$\pm$}l|r@{$\pm$}l||} \hline  \hline
\textbf{Filter} & \multicolumn{6}{c}{\textbf{V407 Vul}} &  
\multicolumn{6}{c}{\textbf{ \rxj\ }}  \\
& \multicolumn{2}{c}{$a_2/a_1$} & 
\multicolumn{2}{c}{$\phi_1$}&\multicolumn{2}{c}{$\phi_2$} & 
\multicolumn{2}{c}{$a_2/a_1$} &
\multicolumn{2}{c}{$\phi_1$}&\multicolumn{2}{c}{$\phi_2$}  \\ \hline
i' & 0.079 & 0.030  & 0.961  & 0.012  & 0.458  & 0.16   & 0.207 & 0.064  &  0.122 & 0.010  & 0.335  & 0.025  \\
g' & 0.053 & 0.014  & 0.970  & 0.005  & 0.388  & 0.08   & 0.157 & 0.025  &  0.117 & 0.004  & 0.338  & 0.012    \\
u' & 0.095 & 0.036  & 0.977  & 0.012  & 0.544  & 0.05   & 0.131 & 0.050  &  0.118 & 0.008  & 0.285  & 0.031   \\[2mm]

r' & 0.024 & 0.015  & 0.960  & 0.020  & 0.361  & 0.099  & 0.202 & 0.009  & 0.1616 & 0.0014 & 0.362  & 0.0034 \\
g' & 0.039 & 0.012  & 0.961  & 0.013  & 0.444  & 0.14   & 0.188 & 0.005  & 0.1659 & 0.0008 & 0.356  & 0.0022  \\
u' & 0.012 & 0.022  & 0.961  & 0.033  & 0.340  & 0.27   & 0.205 & 0.014  & 0.1700 & 0.0022 & 0.342  & 0.0055  \\ \hline \hline
\end{tabular} \\
\caption{First and second harmonic decomposition of the optical light
  curves for V407~Vul and \rxj. $a_1$ and $a_2$ are the semi-amplitudes
  of the first and second harmonics respectively and $\phi_1$ and  $\phi_2$
  their phases of maximum light on Table~\ref{tbl:efe1}'s and \citeauthor{strohmayer05a}'s
  (\citeyear{strohmayer05a})  ephemeris. 
In the case of \rxj, the measurements at the top and bottom come from the WHT
and VLT respectively, hence the marked difference in the uncertainties.\protect\label{tbl:harmo}}
\end{table*}
for \rxj.
We also fitted a sinusoid with frequency fixed to the value derived from the ephemeris of 
Table~\ref{tbl:efe1} at our
observing date to obtain a normalised amplitude of variation and the time (or
equivalently the phase) of the  maximum. The normalised amplitudes,
the phase and the time-shifts relative to the $g'$-band are presented in
Table~\ref{tbl:vulres}.
\begin{table}
\centering
\begin{tabular}{||l|r@{$\pm$}l|r@{$\pm$}l| r@{$\pm$}l||} \hline  \hline
\textbf{Filter} & \multicolumn{2}{c}{\textbf{Semi-amplitude}} &   
\multicolumn{2}{c}{\textbf{$t-t_0$}} & \multicolumn{2}{c}{\textbf{$ 
\phi$}}\\
& \multicolumn{2}{c}{(\%)} & \multicolumn{2}{c}{(s)} & 
\multicolumn{2}{c}{(cycles)}\\ \hline
i' &  3.03  & 0.06 & -4.9  & 1.8   &  0.9612 & 0.0032 \\
g' &  8.47  & 0.09 &  0.0  & 0.9   &  0.9698 & 0.0016 \\
u' &  20.50 & 0.61 &  3.9  & 2.7   &  0.9767 & 0.0047  \\
r' &  4.39  & 0.06 & -0.7  & 1.2   &  0.9596 & 0.0021  \\
g' &  8.70  & 0.07 &  0.0  & 0.7   &  0.9607 & 0.0013 \\
u' &  21.64 & 0.44 &  0.0  & 1.8   &  0.9607 & 0.0033  \\ \hline\hline
\end{tabular} \\
\caption{Results of single harmonic sinusoid fitting for V407~Vul. The 
first three lines show the results for the May 2003 data and the last three 
lines show the results obtained in August 2005. The times mark the
position of the maximum phases and are $T_0= 52782.191666$ for May
2003 $T_0=53612.9483393$ for November 2005. The phases are relative to
the ephemeris of Table~\ref{tbl:efe1}.}
\label{tbl:vulres}
\end{table}
The amplitude increases strongly towards short wavelengths but there is no
observable phase shift with wavelength.  From Table~\ref{tbl:vulres} we
calculate a difference of phase between our two runs of $0.0089 \pm 0.002$. This
could be taken to be as a significant shift in phase, however the uncertainty only
represents the measurement error. When we include the uncertainty of the ephemeris
calculated with Equation~\ref{eq:2} we obtain $0.0089 \pm 0.019$, and therefore
we conclude that there is no significant variation of the phase shift between the optical
and the X-rays between the two epochs of our observations and that the new
ephemeris can be used to extrapolate to later times.  To compare the optical
phases with the X-ray light curves it is important to notice that the absolute
error of the phase due to the ephemeris of Table~\ref{tbl:efe1} is $0.0090$ for May
2003 and $0.019$ for August 2005. 

\subsection{Pulsation astrometry of V407 Vul}

A totally unexpected feature of V407~Vul is that its optical spectrum is
dominated by that of a late G/early K star which cannot fit within a 10-minute
period binary orbit \citep{steeghs2006}. Although this immediately suggests the
IP model in which one expects a main-sequence secondary star \citep{norton04a},
the star shows no radial velocity variations at a level that rules out orbital
periods typical of cataclysmic variable stars
\citep[$\leq 1$day,][]{steeghs2006}.  Alternatives are that this star is a line-of-sight
coincidence (the field is a crowded one), or it could be part of a triple
system with the variable. To discriminate between the latter possibilities we
searched for variations in the position of V407~Vul on its 569 period. The idea
behind this ``pulsation astrometry'' is that although we cannot spatially
resolve the variable and G star components of V407~Vul directly, we can use the
pulsations of the variable to try to detect their separation because their mean
position will move back and forth between the variable and the G star as the
variable brightens and fades.  This method is sensitive to separations well
below the seeing.

\begin{figure*}
\hspace*{\fill}
\includegraphics[width=\columnwidth]{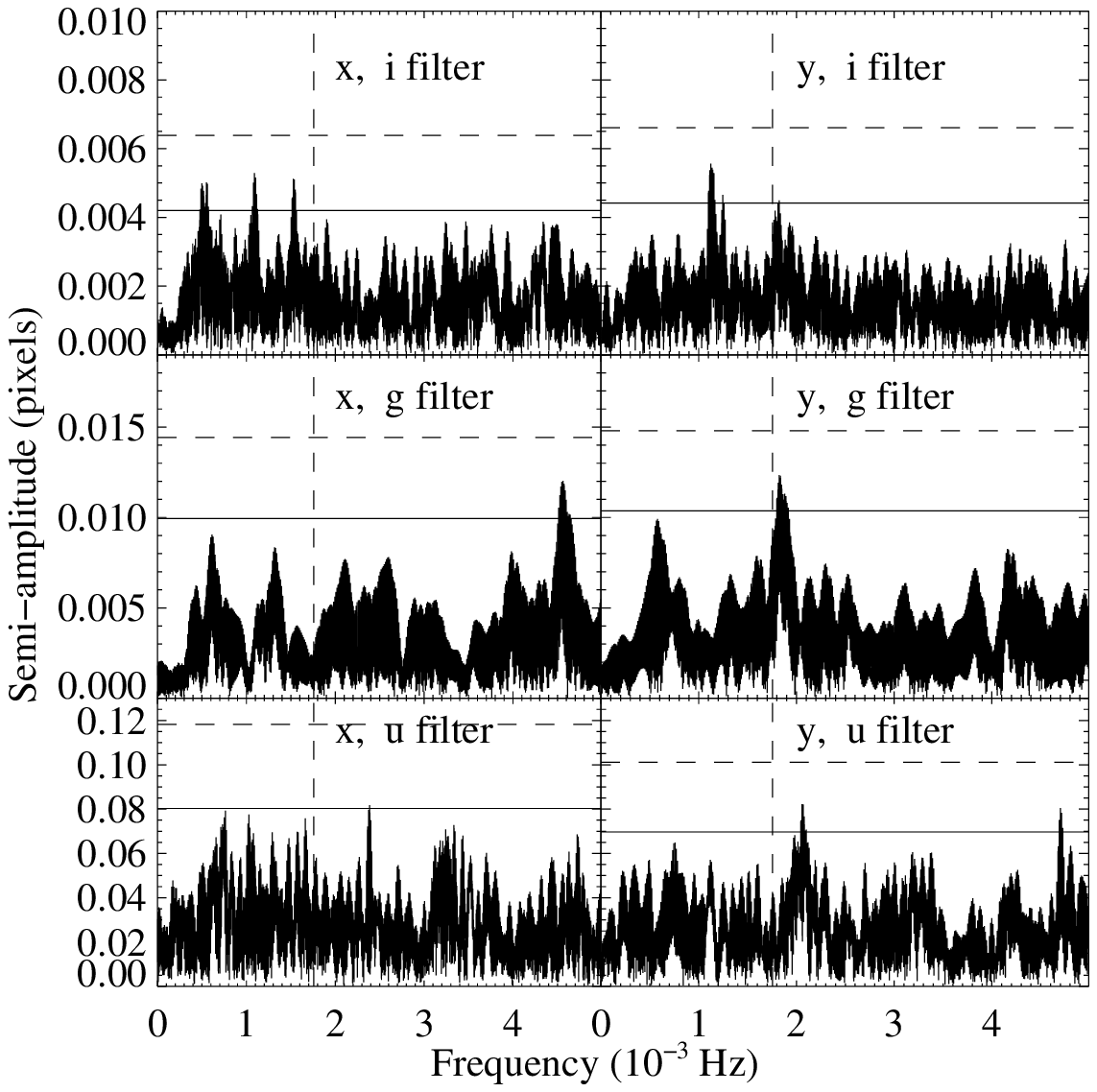}
\hspace*{\fill}
\includegraphics[width=\columnwidth]{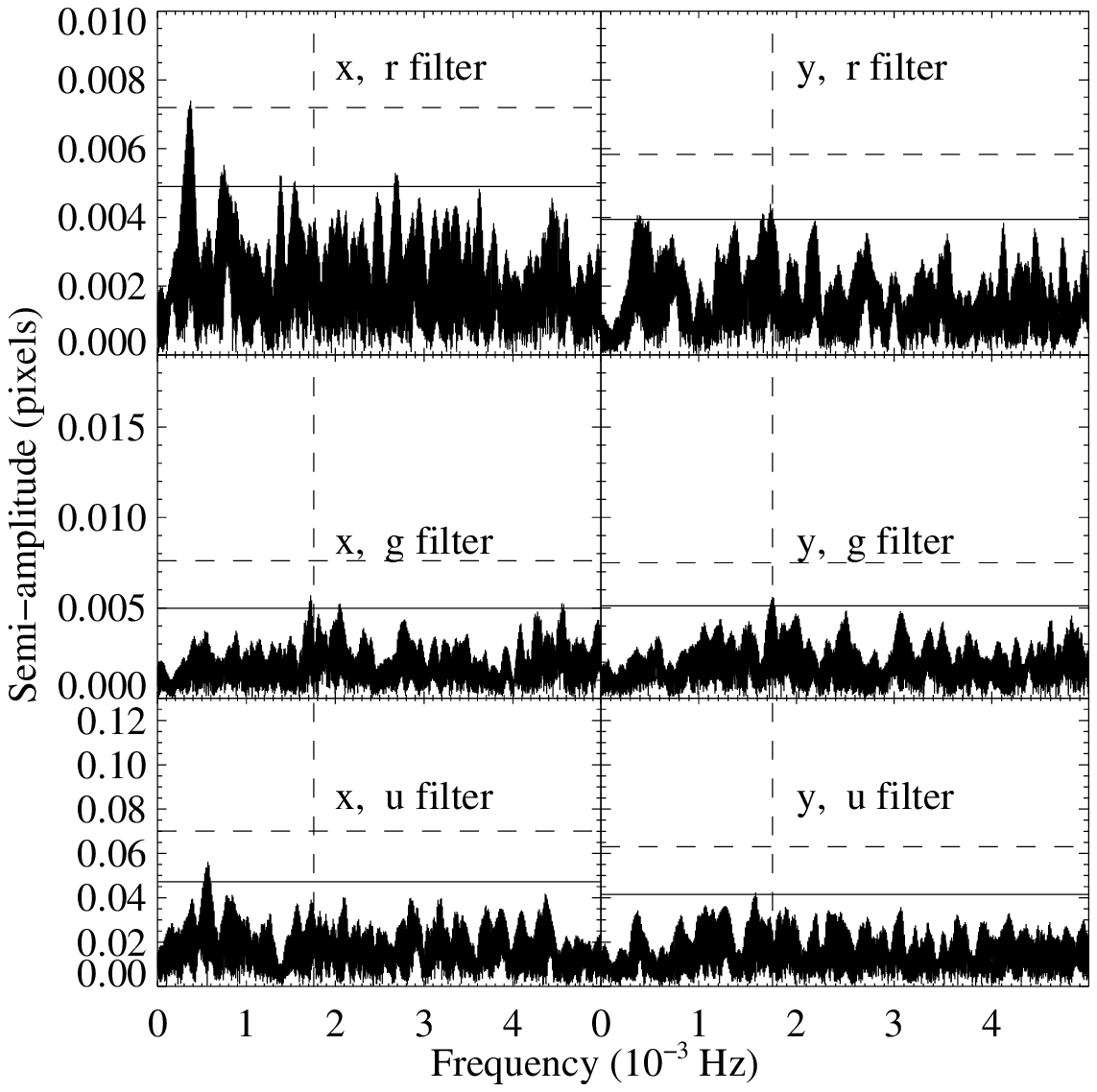}
\hspace*{\fill}
\caption{ Each panel shows the amplitude spectra                    of the x (left) and y
  (right) variation of the position of V407~Vul for the three filters $i'$ or
  $r'$, $g'$, and $u'$ from top to bottom.  The left panel shows the May 2003
  data while the right shows the August 2005 data. The vertical dashed lines show the position of
  the 569 sec period. The solid horizontal lines show the 99.9 percent
  significance level for a known period and the dashed horizontal lines show the
  same level for an arbitrary period.
\label{fig:pos}}
\end{figure*}

We measured the position of V407~Vul relative to nearby stars in the field. 
 We then computed the amplitude of the best-fitting sinusoid over a range of frequencies for both the $x$- and $y$-positions
in each of the three filters as shown in Fig.~\ref{fig:pos}.
We computed false alarm probabilities using Monte Carlo simulations (finding
values that agree with the theoretical values of \citealp{perierro1998}). In
Fig.~\ref{fig:pos} we show the 99.9\% detection threshold for a known period
(horizontal solid lines) and also the 99.9\% detection threshold for an
arbitrary unknown period (dashed lines).  We choose the 99.9\% level because it
corresponds to a detection limit of about ``$3\sigma$''.  Note that the
detection criterion is more stringent when we don't know the period because a
penalty must be paid for searching multiple independent periods
\citep{hornebaliunas1986}. We include this level to show that there are no such
detections of any other periodicities. In the case of V407~Vul we know the
period that we are looking for so it is the lower threshold represented by the
solid lines that applies. As mentioned above, the $g'$ data of the May 2003 run
were partially affected by poor charge transfer in a column close to V407~Vul.
This has more of an effect upon position (especially at the levels we measure
here) than on flux, so for the position measurements we discarded the 50\% of
the $g'$ data where V407~Vul was closest to the column, but as a result reduced
the sensitivity of the $g'$-band data in the left-hand side of the figure.

\begin{figure}
\centering
\includegraphics[width=\columnwidth]{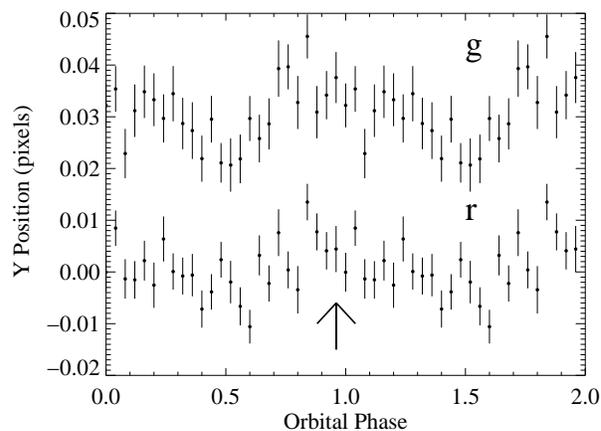}
\caption{Phase folded position variation for $g'$ and $r'$ the datasets that 
show a significant signal. The arrow shows the position of the maximum of 
the flux. \label{fig:pos2}}
\end{figure}

There are detections of a signal at the 99.9\% level in the $y$-position
data in both $r'$ and $g'$ of the August 2005 run. Fig.~\ref{fig:pos2}
shows a phase-folded, binned plot of the $y$-position for these two cases.  The
time of maximum excursion roughly corresponds with the time of maximum light as
expected, and both datasets are consistent with each other in this respect.
However, the signal is tiny, with an amplitude of just $0.005$ pixels or
$0.0015$ arcsec, and so we endevoured to test the reliability of this detection
as far as we were able.  The most obvious problem is that V407~Vul is in a
crowded field and so the position measurements could be affected by other stars.
There are two stars within 1.5 arcsec of V407~Vul that can be seen in
Figure 2 of \citet{ramsay2002a}.  To check how these stars affected our
measurements we first tested whether the detection depended upon the FWHM of the
seeing.  We divided the data in two parts, higher and lower FWHM. The reduction
of data size lead to no detection in either case but the significance of the
peaks was higher in the small FWHM dataset. The reverse would have been expected
had blending with the two nearby stars been the cause.

\begin{figure}
\centering
\includegraphics[width=\columnwidth]{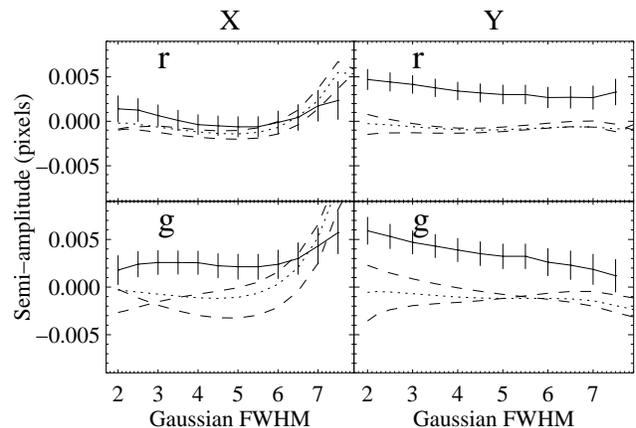}
\caption{The dependence of the amplitude of the variation of the position 
upon the width of the Gaussian used to calculate the position. The data are
plotted as a solid line, the simulations accounting for the known stars are 
plotted as dotted lines (no noise) and dashed lines (with noise). The 
simulations with noise have similar errors to the data but these errors were 
not plotted for clarity. The plate scale is $0.3$'' per pixel.\label{fig:pos3}}
\end{figure}

We measured the centroids by cross-correlation with 2D Gaussians of fixed width.
This allows us to assess the effect of the Gaussian width upon the measured
amplitude. As the FWHM of the Gaussian increases, we expect to see a more
pronounced impact of the nearby stars. Therefore if it is the nearby stars
rather than the G star that are responsible for the variation, we expect an
increase of measured amplitude with Gaussian width. In fact we see the reverse
as Fig.~\ref{fig:pos3}
shows, at least in the $y$-positions for which we have detections. The $x$
positions do show a distinct upturn at large FWHM owing to the much brighter star
5\arcsec\  East of V407~Vul \citep[star B of][]{ramsay2002a} which was positioned to
the left of V407~Vul in our data.

As a final check we carried out simulations of our position
measurements using parameters matching the stars that we could see
nearby V407~Vul, including the two very close ones mentioned above.
This leads to the dotted line in Fig.~\ref{fig:pos3}. In viewing this
figure it must be recognised that the data are not independent and so
to some extent the trends with FWHM can just reflect noise; the dashed
lines in the figure show two simulations of the effect that noise can
have upon the simulated amplitudes. These show that for the
$y$-positions the measured amplitudes are indeed significantly larger
than the simulated values, and provide further confidence in the
reality of the detection.

We conclude, albeit tentatively, that we have detected a change in the spatial
position of V407~Vul that is correlated with its pulsations and that the change
in position is because the G star we see in its spectrum is not exactly
coincident with the variable. We predict that the G star should be below
V407~Vul in our field which roughly corresponds to south of V407~Vul. We obtain
amplitudes of the position variation which we denote by $p$ of $0.00512 \pm
0.0012$ and $0.00514 \pm 0.0010$ pixels for $g'$ and $r'$ respectively.

The value of $p$ is related to the separation on the sky $d$, the fractional
amplitude of the flux variation $a$ as listed in Table~\ref{tbl:vulres} and the
fractional contribution of the G star to the flux at minimum light $f$, 
through the following relation:
\begin{equation}
p =  \left(\frac{a}{a+1}\right) f d .
\label{eq:move}
\end{equation}
Using the measured values for $a$ and $p$ we calculate $ f_{r'} d = 0.0366'' \pm
0.0073$ and $ f_{g'} d = 0.0192'' \pm 0.0046$.  This gives a value of
$f_{g'}/f_{r'} =0.52\pm0.16$.  This is consistent with the spectra of the G star
from which \citet{steeghs2006} estimate that $f_{r'} > 0.85$ and $f_{g'} > 0.6$.
These numbers also match the amplitude of the flux variation whose significant
drop from $u'$ to $g'$ to $r'$ (Table~\ref{tbl:vulres}) can be explained by
dilution of an underlying variable with a constant amplitude with wavelength, as
for \rxj.  If we assume $f_{g'} = 0.7$ we obtain $d \sim 0.027''$; this compares with the upper limit of $0.1"$ set by \citet{steeghs2006}. The distance
to the G star of $1\,\mathrm{kpc}$ estimated by \citet{steeghs2006}, leads to a
minimum separation of $\sim 30\,$AU, equivalent to a period of
$120\,$yr,  and 
means that the G star cannot be the
direct cause of the optical and X-ray pulsations.  Nevertheless it may well be
associated with the system in the form of a hierarchical triple, a point we
return to after we have presented the lightcurves of \rxj. We finish by noting
that our failure to detect anything in the $u'$ band is to be expected. Assuming
typical colours for the G star and a hot spectrum for the variable, we expect
that if $f_g' = 0.7$, then $f_u' = 0.3$. The effect of this reduction in $f$,
which is to make any movement more difficult to detect is in large part offset 
by a factor $2.1$ increase in $a/(1+a)$, but then we are faced with a factor
8 worse sensitivity in the $u'$ band, and the result is that there is no detection in the $u'$ band data.

\subsection{\rxj}

\label{sec:rxj}

We present the phase-folded light curves of \rxj\ using
\citeauthor{strohmayer05a}'s \citeyear{strohmayer05a} ephemeris in the
right-hand panel of Fig.~\ref{fig:v407lc}.

We computed the Lomb-Scargle periodogram to confirm its 321 sec period and we
noticed that the relative strength of the second harmonic is higher than
V407~Vul's, as has already been pointed out by \citet{israel02a}. This is indeed
clear from the non-sinusoidal shape of the light curves in
Fig.~\ref{fig:v407lc}.  The results of the relative strength of the first and
second harmonic and their phases are shown in Table~\ref{tbl:harmo}. The second
harmonic is approximately 15\% of the fundamental and its maximum occurs $0.2$
of a cycle after the maximum of the fundamental. This results in an asymmetry in
the light curve whose rise time is longer than its decline; we discuss its
origin in section~\ref{sec:discu}.
\begin{table}
\centering
\begin{tabular}{||l|r@{$\pm$}l|r@{$\pm$}l| r@{$\pm$}l||} \hline  \hline
\textbf{Filter} & \multicolumn{2}{c}{\textbf{amplitude}} &   
\multicolumn{2}{c}{\textbf{$t-t_0$}} & \multicolumn{2}{c}{\textbf{$ 
\phi$}}\\
& \multicolumn{2}{c}{(\%)} & \multicolumn{2}{c}{(s)} & 
\multicolumn{2}{c}{(cycles)}\\ \hline
i' &  14.77 &0.95 &  1.7 &3.2  & 0.121  &0.010  \\
g' &  13.48 &0.34 &  0.0 &1.3  & 0.116  &0.004  \\
u' &  13.08 &0.66 &  0.5 &2.6  & 0.118  &0.008 \\
\\
r' &  13.54 &0.12  &-1.30  &0.45  &0.1615  &0.0014 \\
g' &  12.74 &0.07  & 0.00  &0.27  &0.1656  &0.0009 \\
u' &  11.90 &0.17  & 1.32  &0.72  &0.1697  &0.0023   \\ \hline\hline
\end{tabular} \\
\caption{Results of single harmonic sinusoid fits to the \rxj\ data. The 
times mark the phases of maximum light and are referenced to two times: $T_0 =52782.895768$ for the May 2003 data and $T_0 = 53702.3368167 $ for 
the November 2005 data.\label{tbl:rxjres2} }
\end{table}

We applied the same method as for V407~Vul to obtain the normalised amplitude of
variation and the time and phase of the maximum. These results are presented in
Table~\ref{tbl:rxjres2}.  In this case the amplitude of the variation decreases
slightly for shorter wavelengths which reinforces the picture that in V407~Vul
the change with wavelength is due to dilution at long wavelengths by light from
the G star.  For \rxj\ the normalised amplitudes of variation are smaller than
the $u'$ band for V407~Vul (which is the least contaminated by the constant
star).  This could be easily explained by the inclination of the plane
of the orbit
and/or differences in temperatures of the stars.

The higher signal-to-noise ratio of the VLT data from November 2005 reveals
a trend with waveband in the phase of the fundamental which is progressively
delayed towards short wavelengths. To test whether the trend is significant, we
carried out an $F$-ratio test comparing two models, one of a constant phase in
the three bands versus one of a linear trend of phase with wavelength, using the
central wavelengths of each band: $3543$, $4770$ and $6222$ \AA.  The
$F$-ratio is the ratio between the $\chi^2$/(number of degrees of freedom) of
one fit divided by the same quantity for the other fit.  We only had three
points so the constant model has two degrees of freedom while the straight-line
fit has just one.  The values of the $\chi^2$ are $10.82$ and $0.068$ for the
constant and straight-line respectively, giving an $F$-ratio of 79.5. This is
significant at the $90\%$ level but not at $95$\%, so, although
suggestive, there is no significant shift.

Table~\ref{tbl:rxjres2} shows that there is a phase difference of
$0.050 \pm 0.004$ between our two runs (May 2003 to November 2005)
where this is the measurement error only. As with V407 Vul, we also
have to add the uncertainty of the ephemeris (see the appendix).  The error of
the difference of phases due to the uncertainty of the ephemeris
calculated using Equation~\ref{eq:2} and the correlation coefficients
given in Table~\ref{tbl:efe2} is $0.013$. So there is a phase
difference between the two runs of $0.050 \pm 0.014$. Therefore there
is marginally significant variation in phase which might mean that
there is a variation of the phase shift between the optical and the
X-rays or, more likely, that the spin up rate is varying.  The
uncertainty in the absolute phase calculated using Equation~\ref{eq:1}
is $0.005$ for the May 2003 data and $0.01$ for the November 2005
data. These are useful to compare the optical phases with the X-ray
phases, and as we shall see next there is a significant phase shift
between the two.

\subsection{The Optical/X-Ray phase shift of \rxj}

The relative phases of the optical and X-ray light curves are an
important constraint upon models.  \citet{israel2003a,israel2004a}
found that optical and X-ray light curves of \rxj\ were in anti-phase
as might be expected for an X-ray emission region facing the secondary
star, contrary, for example, to expectations based upon the direct
impact model.

\begin{figure*}\centering
\includegraphics[width=18cm]{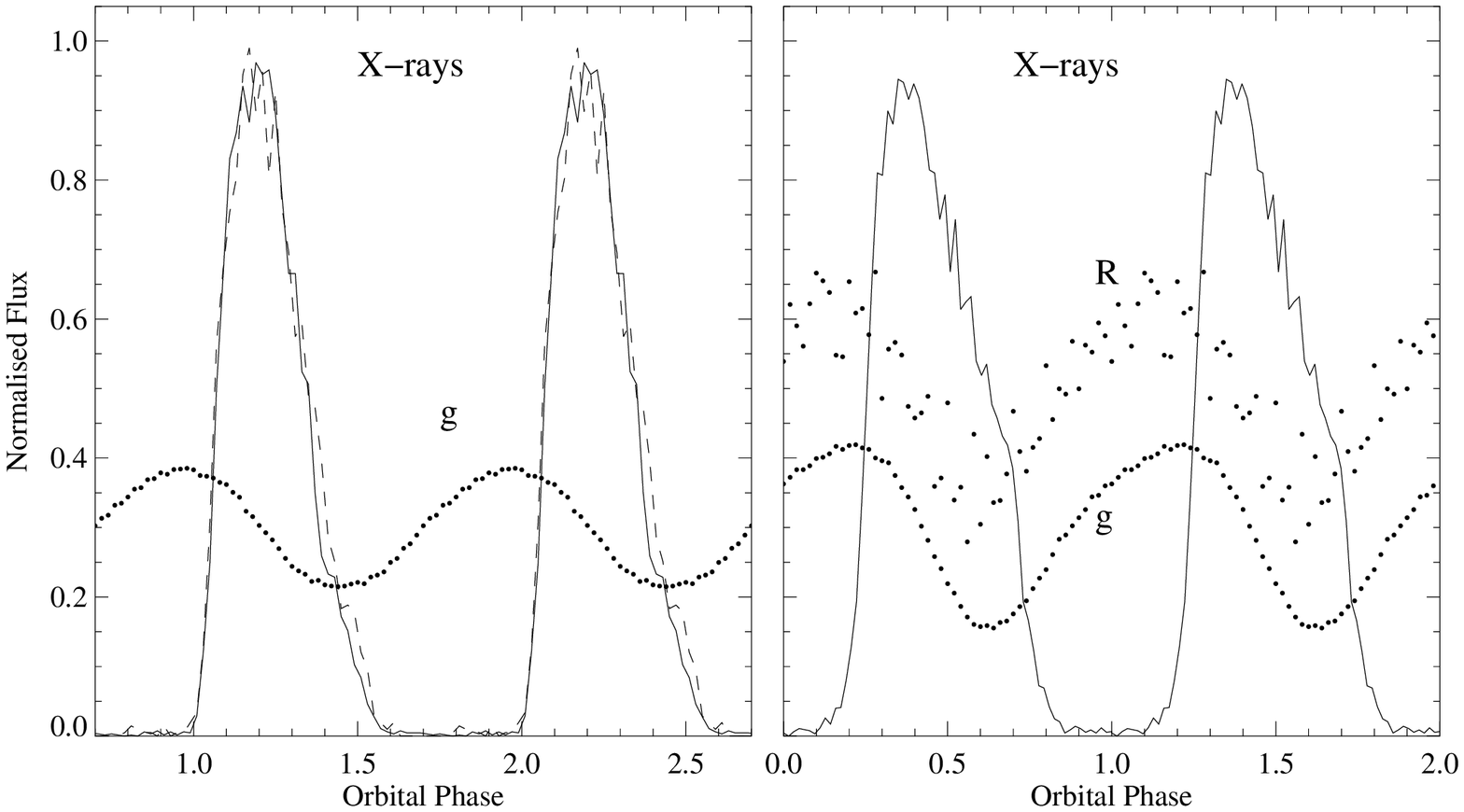}
\caption{ X-Ray/optical phase folded light curve of V407 Vul (left)
  and \rxj\ (right) using the ephemeris of
  \protect{Table~\ref{tbl:efe1}} and \citep{strohmayer05a}
  respectively.  For the X-ray light curves of V407~Vul we re-analysed Chandra
  observations, with the solid line showing data from $19^{\mathrm{th}}$
  February 2003 and the dashed line from $24^{\mathrm{th}}$
  November 2003.  For \rxj, the X-rays were adapted from
  \citet{strohmayer05a}. We overplot the optical $g'$ band showing
  our results. For \rxj\ we also show the VLT/FORS data
  \citep{israel02a} taken in the R filter (top). }
  \label{fig:optxr}
\end{figure*}

In the right-hand panel of Fig.~\ref{fig:optxr} we present the X-ray
and optical light curves folded on \citeauthor{strohmayer05a}'s
(\citeyear{strohmayer05a}) ephemeris. Our phase shift differs from
\citeauthor{israel2003a}'s (\citeyear{israel2003a,israel2004a})
studies by around $0.2$ cycles. To test if this was a genuine change
in the system, we reduced some of the archival $R$-band VLT data from
the $12^{\mathrm{th}}$ of December 2002 used by \citet{israel2004a}.
We reduced these data with the ULTRACAM pipeline and applied the same
methods and time conversions as for the WHT/ULTRACAM data. The results
are also shown in the right-hand panel of Fig.~\ref{fig:optxr} and
agree nicely with our ULTRACAM data.  Clearly the system phase is
stable and there are no problems with the times of either data set.
The difference must be due to the data reduction. We confirmed our
timing results with three different data reduction packages so we
believe that our relative phase is correct and suspect that there is a
problem with \citeauthor{israel2003a}'s
(\citeyear{israel2003a,israel2004a}) values. We were able to confirm
\citeauthor{israel2003a}'s (\citeyear{israel2003a,israel2004a}) X-ray
phase so assume that there is a problem only with the optical timings.
The shift of $0.2$ cycles is about $1$ minute, which is suggestively
close to the $\sim 64$ seconds offset between UTC and TDB. Dr~Israel
was kind enough to confirm that such an error was possible.

\subsection{Flickering}

\begin{figure}
\centering
\includegraphics[width=\columnwidth]{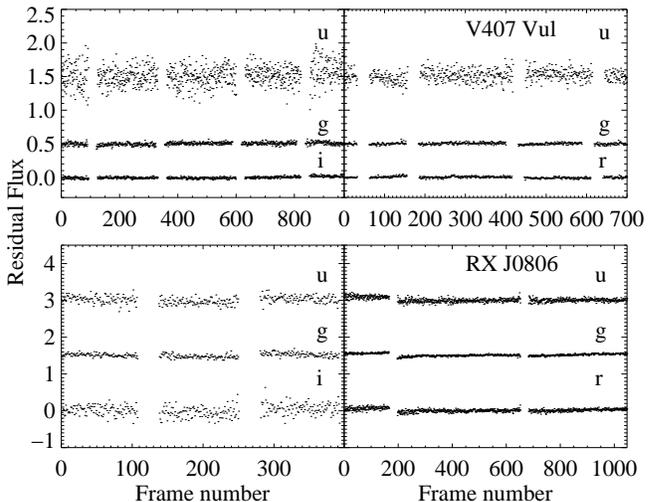}
\caption{V407 Vul (top panel) and \rxj\ (bottom panel) light curves after
  removal of the sinusoidal pulsations. For each object the data from the first
  observing period is in the left and the second in the right. There are no
  significant variations of either source. We inserted gaps between different nights.}
\label{fig:fil2}
\end{figure}

The random stochastic variations known as ``flickering'' are one of the
hallmarks of accreting systems. We therefore looked for any signs of flickering
in our data. A plot of the light curves after removing the sinusoidal variation
is shown in Fig.~\ref{fig:fil2}. 

The light curves are very constant except for long timescale variability of
\rxj\ during the November VLT run. The observations of \rxj\ started at high
airmass, so some of variations seen could be a consequence of extinction, except
that $u'$ does not look much more variable than $g'$ or $r'$. Therefore we
believe that this may be true variability of the source and not an artefact.

In both systems the magnitudes measured in the two observing runs agree well
within the errors.  In the case of V407~Vul we also searched for any flux
variation on longer time scales. We had data from the ``auxiliary port'' of the
WHT taken on the $10^{\mathrm{th}}$ of April 2003 and also Liverpool Telescope data taken on
the $5^{\mathrm{th}}$ of September 2004. The different data sets were all within $10\%$
of each other.

We estimated the variability of these systems by calculating the RMS
of the lightcurves after removing the sinusoidal variations. We
filtered the short-term variations to minimise the photon noise. We
also filtered longer term variations so we could compare our short
runs with a longer run on the cataclysmic variable SS Cyg.  For the
August 2005 V407 Vul run we obtain an RMS variability of 0.7\% in $r'$
and 0.8\% in $g'$. For the November run on \rxj\ we obtain an RMS
variability of 1.6\% in $r'$ and 1.0\% in $g'$. We use these runs as
they have the highest signal-to-noise, nevertheless the variability
still contains a significant component of photon noise. We do not
quote the variability in $u'$ because it is completely dominated by
photon noise.  For comparison, applying the same filtering of the data
to data on the well-known CV SS~Cyg, we obtain an RMS variability of
3.0\% in $r'$ and 5.0\% in $g'$. As mentioned above, the fraction of
the G star in the $g'$ band of V407 Vul is of order 70\%. Therefore
its intrinsic variability is of order 2.5\%, assuming the variations
are not dominated by photon noise. This is a factor of two less than
SS~Cyg and a factor of four if one accounts for SS~Cyg's dilution by
its secondary star \citep{north2002}.  Other cataclysmic variable
stars we looked at are similar to SS~Cyg, so we conclude that the
measured variability in V407 Vul and \rxj\ is much less than in normal
cataclysmic variable stars.

The lack of obvious flickering is a point against accreting models,
although not a conclusive one as there are wide variations in the
amount of flickering shown by definitively accreting binaries, and one
cannot be certain that it should have been detected. It does however
suggest that most of the optical light does not come directly from the
accreting region.

\section{Discussion}
\label{sec:discu}

\subsection{The X-ray versus optical phases}

The correction to the relative X-ray versus optical phase of \rxj\ that we have
identified makes it very similar in this respect to V407~Vul: in each system the
X-ray flux peaks $\sim 0.2$ cycles after the optical flux. This can be added to the
shapes of the light curves as evidence that these two stars are indeed related
systems, as is evident from Fig.~\ref{fig:optxr}.  We will now investigate what
this result implies for the different models.

In the majority of models, the X-rays come from a spot on the primary star which
moves in and out of view as it rotates. The exception is the IP model where the
modulation is the result of the accretion stream flipping from one pole to the
other although it seems unlikely that such a process can really switch off the
X-rays as completely as observed. Less attention has been paid to the optical
pulsations.  Within double degenerate models, these seem likely to originate
from the heated face of the secondary star which would naturally explain their
near-sinusoidal shape and, perhaps, the absence of flickering. Such heating may
be a result of the X-ray emission from the primary star, or the primary star
could simply be hot as a result of compressional heating \citep{bildsten2005}.

\begin{figure*}
\centering
\includegraphics[width=14cm]{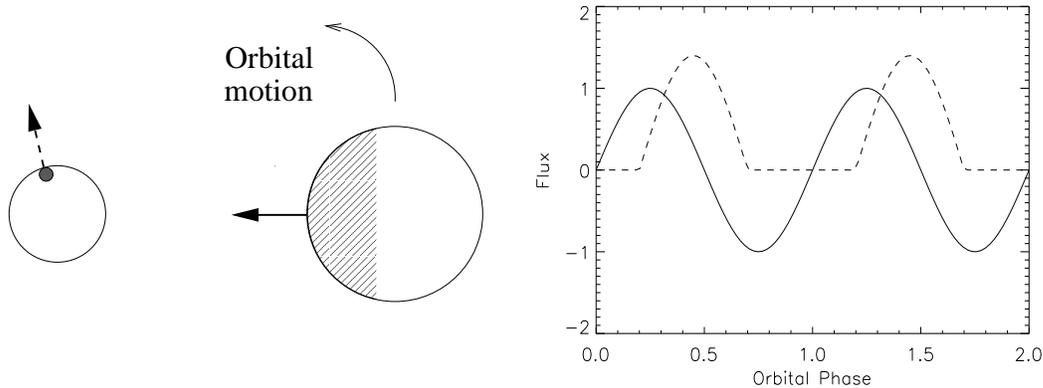}
\caption{In the left-hand panel we show a schematic picture of the primary star
  (left) with an X-ray emitting spot which has a peak of X-ray emission in the
  direction indicated by the dashed arrow. The relative sizes of the two stars
  are drawn to match masses of $M_1 = 0.53\,\msun$ and $M_2 = 0.12\,\msun$ (see
  text). The secondary star, which orbits counter-clockwise, has a heated face
  (shaded) whose peak emission is in the direction of the solid arrow. The
  figure is arranged to give the optical (solid) and X-ray (dashed) light curves
  shown on the right, which have the same relative phasing as both V407~Vul and
  \rxj.}
\label{fig:xray_vs_optical}
\end{figure*}

Assuming that we are correct about the main site of optical emission,
Fig.~\ref{fig:xray_vs_optical} shows the geometrical arrangement that explains
the relative phases of the optical and the X-ray light curves.
In this model the $0.2$ cycles delay of the X-ray peak relative to the optical peak
implies that the X-ray emission spot is rotated $\sim 0.3$ cycles from directly
facing the secondary star, in the direction of the orbit.

This is not the whole story however, because in \rxj\ at least the optical
light-curve is somewhat saw-toothed in shape. As our Fourier decomposition
shows, this is caused by a significant second harmonic that happens to peak at
the same phase as the X-rays, as seen in the difference $\phi_2 - \phi_1 \approx
0.2$ in Table~\ref{tbl:harmo}. The natural explanation for this is that the
X-ray emission spot is also the site of some optical light. If this is a
localised region so that the shape of its light curve can be approximated by a
truncated sinusoid ($f(\theta) = \cos \theta$ for $-\pi/2 < \theta < \pi/2$,
$f(\theta)= 0$ otherwise), then it can produce a second harmonic. It will also
contribute some first harmonic as well, which means that the first harmonic
emission that we see is the combination of contributions from the heated face of
the secondary star and the spot on the primary star. This retards the optical
phase so that the observed $0.2$ cycle shift is an under-estimate of the true
shift between the emission from the heated face and the X-ray emission.

If the optical emission truly can be approximated by the truncated sinusoid,
then for \rxj\ we find that we can fit the phases and harmonic amplitudes if the
X-ray spot leads the heated face of the secondary star by $\sim 0.26$ cycles
(i.e. a little more than $90^\circ$) and the optical emission from the spot on
the primary star peaks at $\sim 75$\% of the amplitude of the emission from the
heated face. While this is probably rather simplistic, it demonstrates that the
simple model illustrated in Fig.~\ref{fig:xray_vs_optical} is capable of
explaining some secondary details of the data. With this decomposition of the
optical light, the X-ray emission site is $\sim 90^\circ$ ahead of the secondary
star and it is then not clear if the X-rays can directly heat the secondary star
or not. V407~Vul has a much weaker second harmonic, and so in this case the spot
is presumably the full $0.3$ cycles or $\sim 110^\circ$ ahead of the secondary
star and cannot see it directly.

The X-ray/optical phase-shifts in V407~Vul and \rxj\ are very naturally
explained by both the direct impact and double-degenerate polar models. The
accretion spot in normal polars is observed to lead the secondary star by of
order $0.1$ to $0.3$ cycles \citep{cropper1988a} and a similar shift is expected
in the direct impact model, depending upon the system parameters
\citep{marsh2004a}. As Fig~\ref{fig:xray_vs_optical} shows, this is exactly what
is required to match the observations.  On the other hand, as far as we can see,
there is no natural explanation for the phase-shift in the unipolar inductor
model for which one would expect anti-phasing, unless there is some as yet
undiscovered mechanism for displacing the magnetic footprint of the secondary
star in advance of its orbit. This is difficult given that the orientation of
the primary star relative to the secondary star changes relatively rapidly in
the unipolar inductor model and so a fixed orientation is hard to contrive.  The
X-ray/optical phase-shift is also a difficulty for \citeauthor{norton04a}'s (\citeyear{norton04a}) IP model
for which they also predict anti-phasing with the optical pulses appearing as
the accretion stream switches to the hidden pole, the X-rays going to zero at
this point. IPs are sufficiently complex that an offset as observed would not
perhaps be that surprising, but in any case there are other more serious
difficulties with the IP model \citep{cropper2004a}.

\begin{figure*}
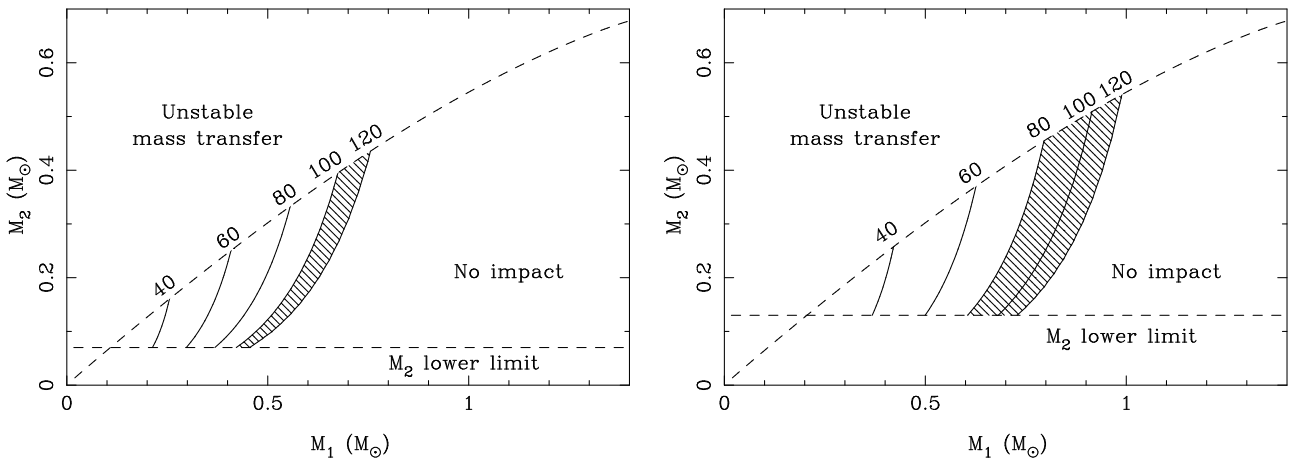

\hspace*{\fill}
\includegraphics[width=6cm,angle=270]{fig8a.ps}
\hspace*{\fill}
\includegraphics[width=6cm,angle=270]{fig8b.ps}
\hspace*{\fill}
\caption{We show the system mass constraints for V407~Vul (left-hand panel) and \rxj\ (right-hand panel) if we assume the direct impact model. The upper dashed line shows the dynamic stability limit. We show contours of equal impact angle. Note that the maximum impact angle is approximately $130^{\circ}$ which corresponds to the transition between the disc and direct-impact accretion. 
  \label{fig:mass}}
\end{figure*}

The direct impact model can be used to predict the phase shape and thus, if it
is true, constrain the masses of the binary components. We define the impact
angle as the angle between the X-ray emission site and the secondary star. We
calculate the impact angle for over a grid of $M_1$ and $M_2$. In
Figure~\ref{fig:mass} we show contours of same impact angle where the shaded
areas represent the probable regions where the systems lie. The uncertainties in
the angle are higher for \rxj\ than for V407~Vul because of the existence of the
second harmonic component. In the same figure we also plotted the dynamic
stability limit (dashed line). We conclude that for V407~Vul $0.4<M_1<0.55\,\msun$ and
$0.08<M_2<0.4\,\msun$ and for \rxj\ $0.6<M_1<0.9\,\msun$ and
$0.12<M_2<0.45\,\msun$. 

\subsection{A limit on the bolometric luminosity of \rxj}
Pursuing the idea of the heated face further leads to a lower limit upon the
bolometric luminosity of \rxj, assuming that the double degenerate models are
correct (unfortunately the G star in V407~Vul's spectrum precludes the same
calculation). The idea is to derive a lower limit on the temperature of the
heated face from the spectrum of the pulsations, which since it is a measure of
the flux from the primary star at a distance equal to the orbital separation,
which is approximately known, gives us a luminosity. A slight complication is
that we do not know for sure whether the X-rays or the primary star's
photosphere is responsible for the irradiation. This ultimately leads us to two
different possible limits. We begin by obtaining the weaker of the two
which applies in the case of X-ray heating, and then consider the revised limit
necessary if the photosphere is responsible for the heating.

\subsubsection{Temperature of the heated face}
\begin{figure*}
\hspace*{\fill}
\includegraphics[width=\columnwidth]{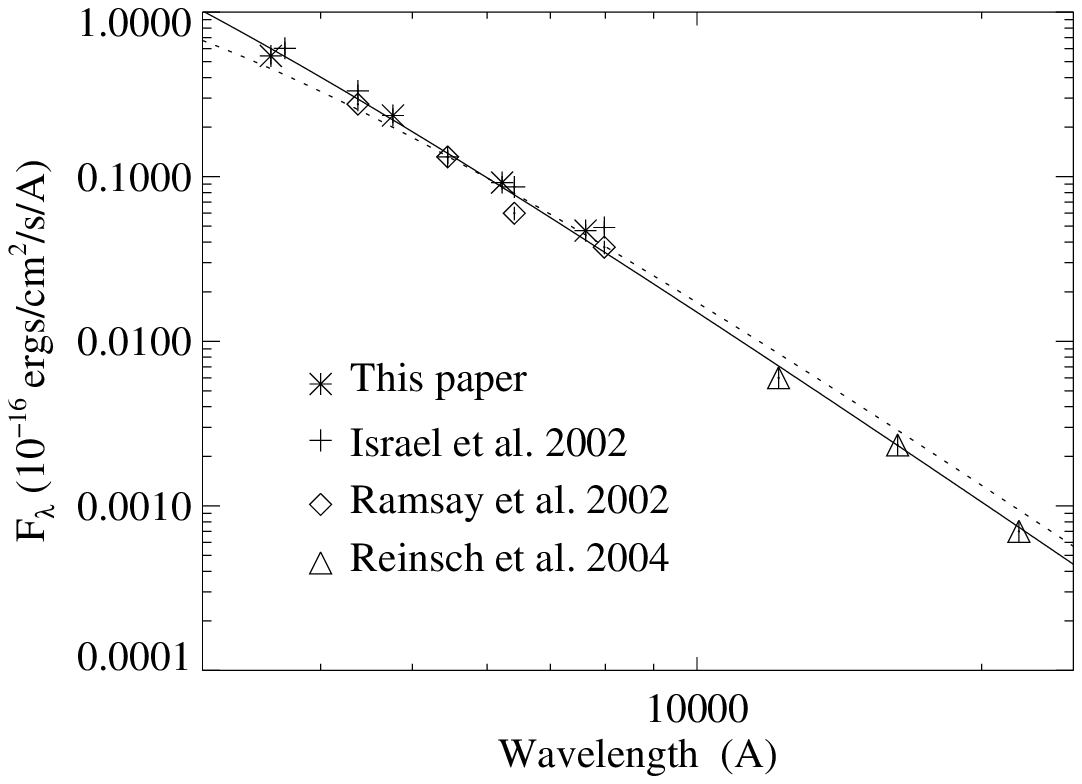}
\hspace*{\fill}
\includegraphics[width=\columnwidth]{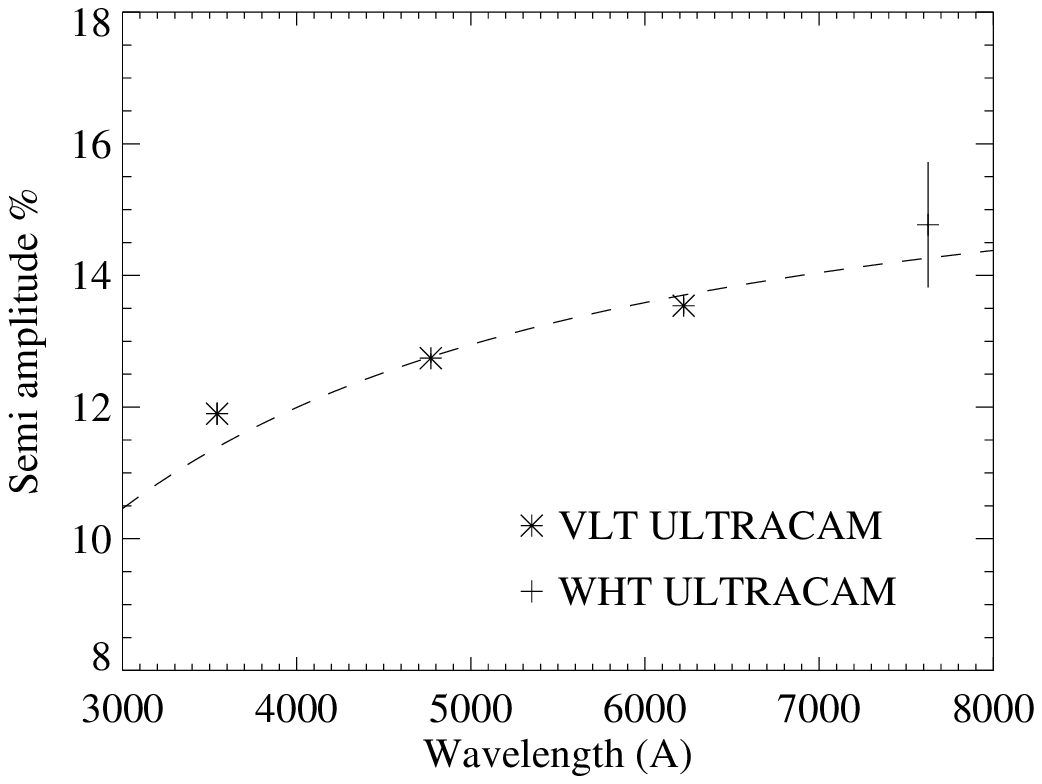}
\hspace*{\fill}
\caption{In the left-hand panel we show the mean fluxes of \rxj\ compared to two
  black-body spectra. One is the best fit ($32$,$400\,\K$, solid), the other has
  the lowest temperature consistent with the data ($T_1 = 18$,$500\,\K$,
  dotted). The right-hand panel shows the pulsation amplitudes that we measure
  compared to the ratio of two black-bodies
  ($B_\lambda(T_\mathrm{irr})/B_\lambda(T_1)$) to place a lower limit upon the
  temperature of the pulsed light $T_\mathrm{irr} > 14$,$800\,\K$; the dashed
  line shows the limiting case.\label{fig:temp}}
\end{figure*}

We first derive a lower limit on the temperature of the spectrum
using our magnitudes and those reported by
\citet{ramsay2002b}, \citet{israel02a} and \citet{reinsch2004a}, as shown in the left-hand panel of
Fig.~\ref{fig:temp}.  
Two black-body spectra are shown, each scaled to give
the minimum $\chi^2$.  One, with temperature $32$,$400\,\K$ (solid line) is the global best
fit, while the other (dotted), with temperature $18$,$500\,\K$, has the minimum
temperature that gives a $\chi^2$ within the 99\% confidence threshold. We take
this to be the minimum possible temperature of \rxj\ given its optical and
infra-red fluxes.  We assume further that this reflects the temperature of the
primary star, $T_1$, since if it is the secondary star, the primary star would
have to be extremely hot to produce a significant reflection effect; there is no
equivalent upper limit as the optical and IR fluxes do not constrain the maximum
temperature at all. Armed with the lower limit of $T_1 > 18$,$500\,\K$, we can
then use the pulsation amplitudes to place a similar lower limit upon the
temperature of the irradiated face of the secondary star $T_\mathrm{irr}$ using the ratio of
black-body spectra, as shown on the right of Fig.~\ref{fig:temp}. Again taking
the 99\% confidence limit, we find that the temperature of the irradiated face
must be at least $T_\mathrm{irr} > 14$,$800\,\K$; this limit rises to
$21$,$700\,\K$ if we use the best-fit value for $T_1$ of $32$,$400\,\K$.  The
lower limit on the temperature of the irradiated face leads directly to a lower
limit on the bolometric luminosity of the primary star since assuming that the irradiation dominates
the intrinsic luminosity of the secondary star we have
\begin{equation}
L_\mathrm{bol} = 4\pi a^2 \sigma T_\mathrm{irr}^4,
\end{equation}
where $a$ is the separation and $\sigma$ is the Stefan-Boltzmann constant.  The
strictest lower limit comes from taking the smallest separation, which
corresponds to the smallest masses for the two component stars. We used $M_1 =
0.53 \,\msun$ and $M_2 = 0.12\,\msun$ which ensure that the secondary star can
fit within its Roche lobe and that mass transfer is stable \citep{marsh2004a}; a
smaller value for $M_1$ could be used if the system is detached, but would be
largely compensated for by the need for a higher value of $M_2$ to avoid mass
transfer. Our masses and the period of $321$ seconds imply a separation of $a =
0.089\,\rsun$ (these values were used to scale Fig.~\ref{fig:xray_vs_optical}).
Scaling from the Sun we therefore find that $L_\mathrm{bol} > 0.34 \,\lsun =
1.3 \times 10^{33}\,\mathrm{ergs}\,\mathrm{s}^{-1}$.

This is already a significant lower limit as it is somewhat higher than, but
consistent with, the X-ray luminosity of $L_X \sim 5 \times
10^{32}\,\mathrm{ergs}\,\mathrm{s}^{-1}$ at $500\,\mathrm{pc}$ distance  estimated by \citet{israel2003a}.
However, we have pushed the temperature to marginally acceptable values. For
instance, the best-fit temperature $T_1 = 32$,$400\,\K$ which leads to
$T_\mathrm{irr} > 21$,$700\,\K$ raises the luminosity limit by a factor of
$4.4$, which hints at a larger distance than \citet{israel2003a} assumed.

Comparing with white dwarfs of similar temperature and mass
\citep{bragaglia1995}, the absolute magnitude of the primary star is bounded by
$M_V < 10.7$.  Given $V = 21.1$ \citep{israel02a}, and assuming that reddening is
negligible, this suggests that $d > 1.1\,\mathrm{kpc}$, although this limit can
be lowered if we adopt a higher mass for the primary star.

These limits apply if it is the X-rays that drive the heating, but it may well
be that it is the photosphere of the primary star itself that is important. The spot position $\sim90^\circ$ ahead of the secondary means that  the spot may not be able to see the secondary star at all. This
leads to a simple but important modification as we now show.

\subsubsection{Heating driven by the primary star's photosphere}

If the primary star's photosphere drives the heating then this sets a relation
between $T_1$ and $T_\rmn{irr}$
\begin{equation}
T_\rmn{irr}^4 = T_2^4 + \left(\frac{R_1}{a}\right)^2 T_1^4,
\end{equation}
where $T_2$ is the temperature of the unheated photosphere of the secondary
star, $R_1$ is the radius of the primary star, $a$ is the orbital separation and
we have assumed that all incident flux is absorbed and for simplicity we do not
try to account for the range of incident angles over the heated face, but just
consider the sub-stellar point. As we said earlier, if $T_2$ is significant, it
is hard to get much of a reflection effect, so we take it to be negligible and
therefore
\begin{equation}
T_1 = \left(\frac{a}{R_1}\right)^{1/2} T_\rmn{irr} .
\end{equation}
The masses adopted above give the smallest ratio of $a/R_1$ leading to $T_1
= 2.58 T_\mathrm{irr}$. This equation was used to bootstrap from the lower limit
on $T_\mathrm{irr}$ to a new lower limit of $T_1$, which was then used to place
a new lower limit on $T_\mathrm{irr}$ using the procedure of the previous
section. We obtained updated limits as follows: $T_\mathrm{irr} > 34$,$800\,\K$
and $T_1 > 90$,$000\,\K$. The heated face temperature rises by a factor of
$2.4$, and so the lower limit on the bolometric luminosity rises by a factor of $33$
to $L > 10 \,\lsun = 4.0 \times 10^{34} \,\mathrm{ergs} \, \, \mathrm{s}^{-1}$.
Again comparing with white dwarfs of similar temperature and mass
\citep{bragaglia1995}, the absolute magnitude of the primary star is bounded by
$M_V < 8.0$. Given $V = 21.1$ \cite{israel02a}, and assuming that reddening is
negligible, we must have $d > 4.2\,\mathrm{kpc}$.  This would place \rxj\ more
than $2.5\,\mathrm{kpc}$ out of the plane, and it would possibly  be a halo object.
We note that a halo-like transverse velocity of $200\,\kms$ and our distance
limit imply a proper motion $< 0.01$ arcseconds/yr, below the limit of $0.02$
arcseconds per years placed by \citet{israel02a}.

Our distance limits do not discriminate between accretion models
which work best for large distances, in excess of $4\,\rmn{kpc}$
\citep{bildsten2005,DAntona2006} and the unipolar inductor model which works
well for $d < 1\,\rmn{kpc}$ \citep{dallosso2006a,dallosso2006b}. However, they do suggest
that UV observations may have a value in tightening the lower
limits upon temperatures and hence the distance.

\subsection{Direct impact or polar?}

We have lumped the accreting double-degenerate models, direct impact
and polar together as ``accretion models'', as we think they provide
equally good explanations for our data. For V407~Vul the
double-degenerate polar model suggested by \cite{cropper1998a} was
discarded when no polarisation was found \citep{ramsay2002a}. In the
case of \rxj, \cite{reinsch2004a} have claimed a detection of
circular polarisation but at a low level given the faintness of the
target (0.5\%) that needs confirmation. However, we think that the
polar model may have been written off prematurely as there are some
very high-field polars which show very little polarisation (AR~UMa,
\citeauthor{schmidt1996a} \citeyear{schmidt1996a}; V884~Her,
\citeauthor{szkody1995a} \citeyear{szkody1995a},
\citeauthor{schmidt2001a} \citeyear{schmidt2001a}) and strong soft
X-ray components, very much like V407~Vul and \rxj. It has been
suggested that this is because the shocks are buried in these systems,
due to the high accretion rates, rather as \cite{marsh2002a} suggested
for the direct impact model. Polars show stronger optical line
emission than either V407~Vul or \rxj, but this is not a strong
argument against the polar model since the systems, if they are double
degenerates, would be helium-rich and very compact, and so different
from normal CVs.

\subsection{Period changes in V407~Vul and the G star}

We have shown that the G star does not play a direct role in the variability of
V407~Vul but it could be gravitationally bound to the variable, in which case it
may cause an apparent period change through variable light travel time effects.
How significant could this be?  Assuming that the G star has mass $M$, then the
maximum acceleration of the binary along the line of sight is $\sim G_c M / a^2$
where $a$ is the separation of the binary and the G star. The subscript $c$ in
the gravitational constant is to avoid confusion with the G star. This leads to
a quadratic term in the usual $T_0 + P E + C E^2$ ephemeris equal to
\begin{equation}
C = \frac{G_c M P^2}{2 c a^2},
\end{equation}
where $c$ is the speed of light and  $P$ the orbital period. This leads
to an apparent rate of period change given by 
\begin{equation}
\dot{P} = \frac{G_c M P}{c a^2}.
\end{equation}
Taking $a$ to be comparable to the projected separation at
$1\,\rmn{kpc}$ of $27\,$AU, and $M = 1 \,\msun$ gives
$|\dot{P}|_\rmn{max} \sim 1.6 \times 10^{-11}\,\rmn{s}/\rmn{s}$.  This
is about 5 times larger than the observed value \citep{strohmayer05a}
and thus we conclude that the G star has the potential to have a
significant effect upon the rate of period change measured in this
system. This adds an extra uncertainty that may allow both the
unipolar inductor \citep{marsh2005a,dallosso2006a,dallosso2006b} and
accreting models \citep{DAntona2006} to match this system more easily
than has been the case to date. Continued observations in order to
place limits upon or detect a relative proper motion between the
variable and the G star would be of interest for testing the triple
star model. We estimate the orbital velocity of the G star to be $\sim
3 \mathrm{km}\mathrm{s}^{-1}$, which is perhaps detectable given a
long enough period of time.

\section{Conclusion}

We have presented optical photometry of V407~Vul and \rxj\ in $i'$,
$r'$, $g'$ and $u'$ bands taken with the high-speed CCD camera
ULTRACAM.  For V407~Vul we have a hint of detection of a third
component in the system at $0.027''$ from the variable. We believe
this to be the G star that is seen in the spectrum of V407~Vul,
which therefore cannot be the secondary star of the variable.  We
cannot distinguish whether it is a line-of-sight coincidence or a triple
system.

For \rxj\ we find a new phasing of the X-ray and optical data which
renders it indistinguishable from V407~Vul with the optical pulses
$0.2$ cycles ahead of the X-ray pulses. The offsets are naturally
produced by double-degenerate accreting models of the systems, both
polar and direct impact, but seem hard to reconcile with the unipolar
inductor and intermediate polar models.  The optical light curves of
\rxj\ are non-sinusoidal and a Fourier decomposition shows that there
is likely a contribution to the optical light from the same site as
produces the X-rays.

On the assumption that the optical pulses of \rxj\ are the result of
irradiation of the secondary star within a double degenerate binary,
and using the relative constancy of the fractional pulsation amplitude
with wavelength, we place a lower limit on the distance to the system
of $> 1.1\,\mathrm{kpc}$. If it is the photosphere of the accretor
rather than the X-ray site that is responsible for the heating, then
this limit rises to $d > 4.2\,\mathrm{kpc}$. Space ultraviolet
observations are the best hope for strengthening these constraints.

Finally we remark that both the polar and direct impact models provide
equally good explanations of our observations and that there are high
magnetic field polars that show similar properties to V407~Vul and
\rxj\, i.e. very soft X-ray spectra and low polarisation.

\section{Acknowledgements}
We thank Gavin Ramsay for his help and advice concerning the X-ray
timings of V407~Vul. We thank Tod Strohmayer and Gianluca Israel for
answering our queries. We thank Christopher Deloye for useful
discussions.  SCC Barros is supported by Funda\c{c}\~{a}o para a
Ci\^{e}ncia e Tecnologia e Fundo Social Europeu no \^{a}mbito do III
Quadro Comunit\'{a}rio de Apoio.  TRM acknowledges the support of a
PPARC Senior Research Fellowship. DS acknowledges support of a
Smithsonian Astrophysical Observatory Clay Fellowship. PJG and GR are
supported by NWO VIDI grant 639.042.201 and GN by NWO VENI grant
639.041.405. Our data were acquired on the $4.2$m WHT of the Isaac
Newton Group of Telescopes, La~Palma, on the Liverpool Telescope and
on the ESO VLT at Paranal in Chile. The ESO proposal was 076.D-0228.
We also made use of the ESO Paranal, NASA ADS and SIMBAD data
archives.

\bibliographystyle{mn2e}
\bibliography{barros}

\appendix
\section[]{Ephemerides}

\subsubsection{ V407~Vul's ephemeris} 

As we mentioned in Section~\ref{sec:efi}, in order to calculate the
uncertainty in the published ephemerides we need the covariance terms
of the fitted coefficients that are not given in any published work.
Therefore we had to digitise and fit the X-ray data in order to obtain
a timing solution whose uncertainties we could compare with our data.
To digitise the data we applied the Linux utilities \emph{pstoedit}
and \emph{xfig} to the PostScript figures from the published papers to
obtain the coordinates of the points and their error bars. Such a
process can at best match the original data, and can potentially
degrade it, but in this case the precision of the PostScript data is
good enough that it has no measurable effect; we were able to confirm
our numbers directly in one case after Dr~Ramsay kindly sent us his
data.  For V407~Vul's ephemeris we digitised the bottom panel of
Figure~6 from \cite{strohmayer04a} and Figure~1 from the recently
published ephemeris of \cite{ramsay2006} that extends the
ephemeris to April 2004. These figures show the residuals in phase
relative to a given timing solution. In each case we applied the given
timing solution to the observation times and added the phase residuals
to obtain the phases as a function of time.  We then fitted a timing
solution similar to the one used by \cite{strohmayer04a}, i.e.
$\phi(t)= \phi_{0}+\nu(t-t_0) + \dot{\nu}(t-t_0)^2 /2 $. We included
an extra term ($\phi_{0}$) because we fixed $t_0$ to be the same as
\cite{strohmayer04a}, so this value is no longer arbitrary.  We obtain
the same fitted parameters as \cite{strohmayer04a} but slightly
different parameters to those of \cite{ramsay2006}.  For reasons we
shall explain, it is our fit to the data of
\cite{ramsay2006} that is given in Table~\ref{tbl:efe1}. 
\begin{table}
\centering
\begin{tabular}{||l|l||} \hline  \hline
$t_{0}\, (\mathrm{TDB})$ &$49257.533373137$	  \\
$\phi_{0}$ &0.003(30)   \\ 
$\nu \,(\mathrm{Hz})  $ &$0.00175624626(39)$   \\ 
$\dot{\nu} \, (\mathrm{Hz\, s^{-1}})   $ &$9.9  (1.9) \times 10^{-18}  $  \\ 
$r(\phi_{0},\nu)$ &-0.92074289   \\ 
$r(\phi_{0},\dot{\nu})$ & 0.86174740  \\ 
$r(\nu,\dot{\nu})$ &-0.98817908   \\ \hline\hline
\end{tabular} \\
\caption{Ephemeris of V407~Vul derived from the data of  \protect\cite{ramsay2006}.The uncertainties of the parameters are given within parentheses.  We also give the correlation coefficients for the fitted parameters.}
\label{tbl:efe1}
\end{table}
This corresponds to $\dot{P} =-3.21(61) \times 10^{-12}\, \mathrm{s
  \, s^{-1}}$ which can be compared with the value $\dot{P} =
-3.31(09) \times 10^{-12}\, \mathrm{s \, s^{-1}}$ from
\citet{ramsay2006}.  Our uncertainty is six times larger than that of
\cite{ramsay2006} and the values are slightly different because we allowed more freedom
in the fit. We think our fit is the correct one because there is no
reason that the fit has to have zero value and gradient at $t = 0$, as
was effectively assumed by \cite{ramsay2006}, who fitted only a parabolic
term.

Comparing the two ephemerides we found that the value of the frequency
derivative was not consistent. Moreover if we calculate the phase of
the maximum of our observation with the ephemeris from Strohmayer we
obtain $0.8170 \pm 0.0016$ for the May 2003 and $0.7328 \pm 0.0013$
for August 2005 (different from $~0.97$ of Table~\ref{tbl:rxjres2}
from \citealp{ramsay2006}). This suggests that the two ephemerides
have different zero points contrary to what was stated in the
respective papers. Next we compared the phases predicted by the two
ephemerides expecting to see a constant offset between the two.
Instead we found a drift between one ephemeris and the other. The
phase difference started at approximately zero for the first
observation with ROSAT (so indeed the two ephemerides had exactly the
same zero point) but were 0.15 cycles apart for the last Chandra
observation. Therefore we re-analysed the ROSAT data from the
$30^{\mathrm{th}}$ April 1996 and phase-folded it on
\citeauthor{strohmayer04a}'s (\citeyear{strohmayer04a}) ephemeris, but
obtained the same phasing as \cite{ramsay2006}. We can only obtain the
same phasing as \cite{strohmayer04a} if we do not apply the UTC to TT
correction. In the ROSAT documentation it says that the times are in
UTC\footnote{http://wave.xray.mpe.mpg.de/rosat/doc/}. The error in the
ROSAT times causes \cite{strohmayer04a} to underestimate the rate of
spin-up in V407~Vul, and this is why his frequency derivative is lower
than that of \cite{ramsay2006}. 

To be sure of the correct phasing between the X-ray and optical light
curves of V407~Vul we used Table~\ref{tbl:efe1}'s ephemeris and
applied it to Chandra data taken on $19^{\mathrm{th}}$ of February
2003 and $24^{\mathrm{th}}$ November 2003, which were taken before and
after our May 2003 observation. We obtained the same relative phasing
of the optical and X-rays at each epoch. When we use the ephemeris of
\cite{ramsay2006}, the two Chandra X-ray light curves are almost
perfectly aligned (Figure~\ref{fig:optxr}), but if we use
\citeauthor{strohmayer04a}'s ephemeris, there is a distinct shift
between them. We take this as further evidence of a problem with
\citeauthor{strohmayer04a}'s ephemeris.

\begin{figure}
\centering
\includegraphics[width=\columnwidth]{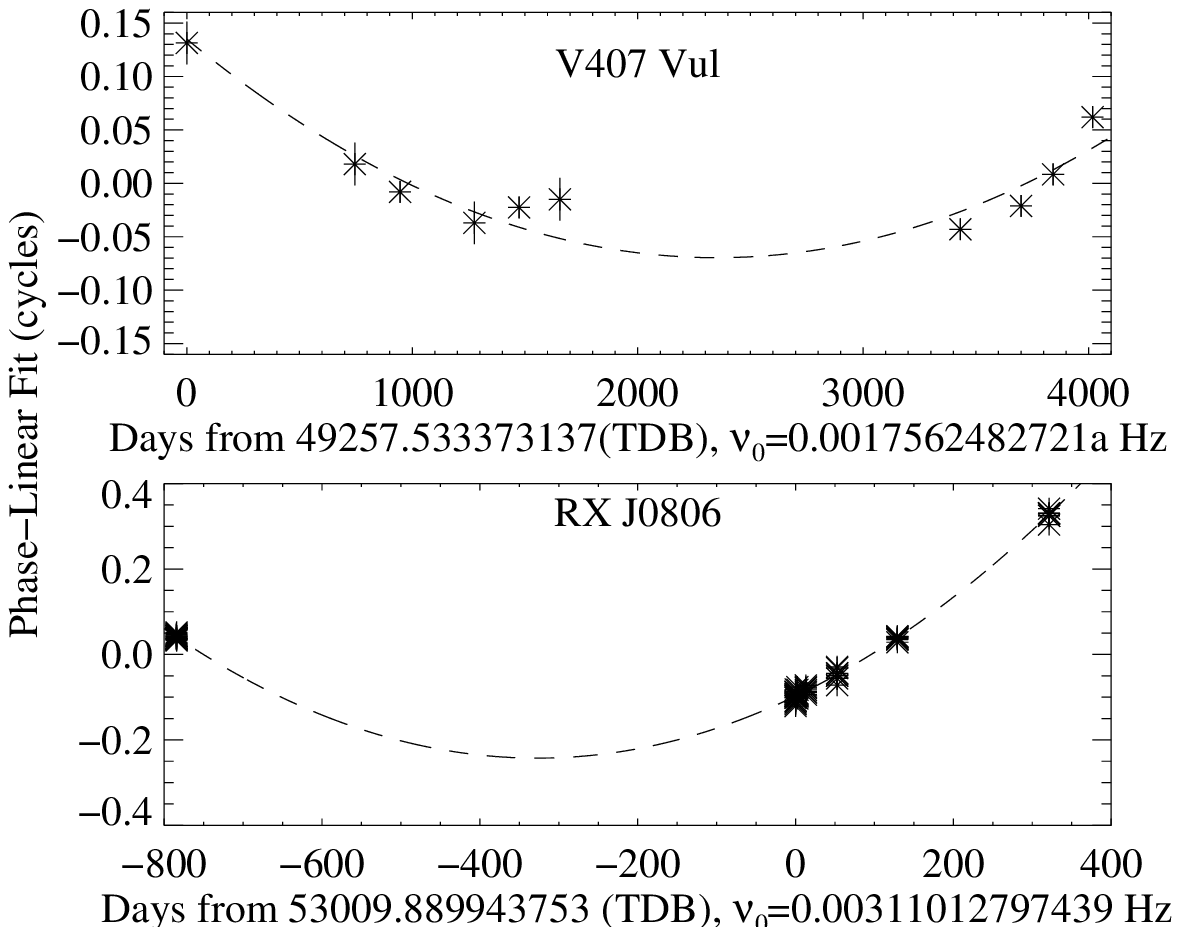}
\caption{Phase residuals of V407 Vul (top panel) and \rxj\ (bottom
  panel) after subtraction of constant frequency models. The dashed line shows our parabola ephemeris subtracted the linear fit. }
\label{fig:efis}
\end{figure}

To conclude, we used the ephemeris that resulted from refitting the
data \cite{ramsay2006} to give the ephemeris listed in
Table~\ref{tbl:efe1}. In the top panel of Figure~\ref{fig:efis} we
show the residuals of our fitted phases for V407~Vul after
removal of a constant frequency model with $\nu_0= 0.0017562482721063\,
\mathrm{Hz}$. We also show the fitted parabola minus the linear fit.

For V407~Vul we performed an $F$-ratio test for the parabola versus the
linear fit. The $\chi^2$ value of the parabola is $23.4$ and of the
linear fit is $115$, we have $10$ points so we obtain an
$F$-ratio=$4.33$ which is significant at the $95\%$ confidence level but not
at $99\%$.

\subsubsection{\rxj's ephemeris}

As was explained above in the case of V407~Vul, in order calculate the
uncertainties in phase due to the ephemeris we need to know the
covariance terms of the ephemeris.  Therefore we applied the same
method as before and digitised and fitted the data of Figure~7 of
\cite{strohmayer05a}. We obtained the same fit coefficients as
\cite{strohmayer05a} so our digitisation does not cause loss of
information; in this case no ROSAT data were involved. We also obtain
the covariance terms. Our fitted parameters are given in
Table~\ref{tbl:efe2}. We show the phase residuals after subtracting a
constant frequency model ($\nu_0=0.0031101279743869\, \mathrm{Hz} )$
in the bottom panel of Figure~\ref{fig:efis}. We applied an $F$-ratio
test to the \rxj\ data. This time there were $69$ points and we
obtained a $\chi^2$ of $54.9$ for the parabolic fit and $10380$ for
the linear fit. This gives an $F$-ratio of $186$, significant at
the $99.99\%$ confidence level.

\begin{table}
\centering
\begin{tabular}{||l|l||} \hline  \hline
$t_{0}\, (\mathrm{TDB})  $ &$53009.889943753 $	  \\
$\phi_{0}$ &0.0003(14)  \\ 
$\nu \,(\mathrm{Hz})  $ &$ 0.00311013824(10)   $   \\ 
$\dot{\nu}  \, (\mathrm{Hz\, s^{-1}})    $ &$ 3.63(0.04)\times 10^{-16}  $  \\ 
$r(\phi_{0},\nu)$ &-0.48041115   \\ 
$r(\phi_{0},\dot{\nu})$ & -0.61096603  \\ 
$r(\nu,\dot{\nu})$ & 0.94898169  \\ \hline\hline
\end{tabular} \\
\caption{Ephemeris of \rxj\ derived from Figure 7 of  \protect\cite{strohmayer05a} . The
  uncertainties of the parameter's are given within brackets.  
We also give the correlation coefficients for the fitted parameters.}
\label{tbl:efe2}
\end{table}

\subsubsection{Uncertainties on phases from the ephemerides}
To calculate the uncertainties in the absolute phases due the uncertainties in
the ephemeris we used the relation:
\begin{eqnarray}
\sigma_\phi^2 &= &\sigma^2_{\phi_0}+(t-t_0)^2\sigma^2_\nu+(t-t_0)^4\sigma^2_{\dot\nu}/4 + \nonumber\\
&&2(t-t_0)C_{\phi_0\nu}+(t-t_0)^3C_{\nu\dot\nu}+(t-t_0)^2C_{\dot\nu\phi_0}
\label{eq:1}
\end{eqnarray}
where $C_{XY}$ is the covariance of $X$ and $Y$ and can be written $C_{XY} =
\sigma_X\sigma_Y r(X,Y)$. We give the correlation coefficients  $r(X,Y)$ in Table~\ref{tbl:efe1} and Table~\ref{tbl:efe2}.

For the phase difference between two epochs $\Delta\phi=\phi(t_2)-\phi(t_1)$ one
cannot simply combine in quadrature the uncertainties on the absolute phases at
each epoch because the same coefficients are used in each case. (This is most
easily seen by considering the case of two identical epochs for which the
uncertainty in the phase difference must be zero.) Instead one must use
the following relation:
\begin{eqnarray}
\sigma^2_{\Delta\phi}&= &(t_2-t_1)^2\sigma^2_\nu+ [ (t_2-t_0)^2-(t_1-t_0)^2]^2\sigma^2_{\dot\nu}/4+ \nonumber\\
&&(t_2-t_1)[ (t_2-t_0)^2-(t_1-t_0)^2]C_{\nu\dot\nu}.
\label{eq:2}
\end{eqnarray}
We used this to calculate the uncertainties on the phase differences in
sections~\ref{sec:v407vul} and \ref{sec:rxj}.

\label{lastpage}

\end{document}